\documentclass[12pt]{iopart}
\usepackage{graphicx}
\usepackage[latin1]{inputenc}
\usepackage[english]{babel}

\begin {document}

\title{Some aspects on the feasibility of satellite quantum key distribution}

\author{C. Bonato, A. Tomaello, V. Da Deppo, G. Naletto, P. Villoresi}
\address{Department of Information Engineering, University of Padova, Italy}

\ead{bonatocr@dei.unipd.it}

\begin {abstract}
In this paper we address some aspects on the feasibility of satellite quantum communication which we believe are still not well understood. We focus on the techniques to get a high enough SNR (in particular in the uplink) and to implement a polarization-preserving channel...
\end{abstract}

\section {Introduction}
For the last decades, a strong research effort has been devoted to study how quantum effects may be employed to manipulate and transmit information, in what is called Quantum Information Processing \cite{nielsenChuang, physQuantInfo, jaegerBook}. These research activities lead to new information-processing protocols with no classical counterpart, like quantum key distribution \cite{gisinReviewQKD, sashaBook, reviewQKDLo}, quantum teleportation \cite{teleportBennettPRL93} or quantum computing \cite{quantumComputingLaflamme}.
Quantum Key distribution, in particular, is on its way from research laboratories into the real world. Fiber and free-space links have been realized linking nodes at larger and larger distances \cite {yamamotoNPhot07, ursin07} with higher and higher key generation rates. Network structures have also been demonstrated recently, for example the DARPA network in Boston \cite {sashaBook} and the SECOQC network in Vienna \cite{secoqc}.

However, current fiber and free-space links cannot implement a real global-scale quantum key distribution system. Fiber links have the advantage that the photon transfer is scarcely affected by external conditions, like background light, weather or environmental obstructions. On the other hand the extension of fiber links beyond few hundred kilometers is problematic, due to attenuation and polarization-preservation issues.
Terrestrial free-space links show some advantages: the atmosphere provides low absorption and is essentially non-birefringent, allowing almost unperturbed propagation of polarization states. On the other hand, the optical mode is not confined in a waveguide, so they are extremely sensitive to the external environment: objects interposed in the line of sight, beam distortion induced by atmospheric turbulence, bad weather conditions and aerosols.

A solution to this problem can be the use of Space and satellite technology. Space-based links can potentially lead to global-scale quantum networking since they allow a much larger propagation distance of photonic qubits compared to present terrestrial links. This is mainly due to the fact that most of the propagation path is in empty space, with no absorption and turbulence-induced beam spreading, and only a small fraction of the path (corresponding to less than 10 km) is in atmosphere. However many technical problems must be overcome in order to realize a working quantum communication link between Earth and Space. Geostationary satellites are too distant to implement a single photon link, therefore fast-moving low-orbit satellites (LEO orbit, from 500 to 2000 km above Earth surface) must be employed.

Several proof-of-principle experiments in this direction have been performed recently. In 2005 C.-Z. Peng and coworkers reported the first distribution of entangled photon pairs over 13 km, beyond the effective thickness of the aerosphere \cite {peng05}. This was a first significant step towards satellite-based global quantum communication, since it showed that entanglement can survive after propagating through the noisy atmosphere.

In 2007 two experiments were carried out at Canary islands by a European collaboration. Entanglement-based \cite {ursin07} and decoy-state \cite {tobias07} quantum key distribution was realized on a 144 km free-space path, linking La Palma with Tenerife. For these experiments the Optical Ground Station of the European Space Agency, developed for standard optical communication between satellites and Earth, was adapted for quantum communication. It is important to highlight that the twin-photon source was able to achieve coincidence production rates and entanglement visibility sufficient to bridge the attenuation expected for satellite-to-ground quantum channels.

In a successive experiment, the feasibility of single-photon exchange for a down-link between a LEO satellite and an optical ground station (Matera Laser Ranging Observatory, in the South of Italy) was experimentally demonstrated \cite {villoresiIOP08}. The researchers exploited the retroreflection of a weak laser pulse from a geodetic satellite covered with corner-cubes (Ajisai, orbiting at around 1400 km) to simulate a single photon source on a satellite. They showed that, implementing a strong filtering in the spatial, spectral and temporal domain the emitted photons can be recognized against a very strong background.

In this paper, we present a detailed analysis of the feasibility of satellite-based quantum communication which we believe have not yet been adequately discussed in the literature. In particular, we concentrate on two issues that were identified as crucial by the experiment performed at Matera Observatory \cite{villoresiIOP08}: the possibility of a good signal-to-noise ratio (SNR) and the polarization mainteinance in the link. As regards the SNR we will refine the models already presented in the literature by introducing a detailed analysis of the effect of atmospheric turbulence and of the background stray-light in the case of a ground-to-satellite uplink. We will then discuss some filtering techniques to improve the SNR reducing the noise level; in particular we will analyse in detail the possibility of high-accuracy temporal filtering. Finally, as long as polarization control is concerned, we will discuss and compare different strategies to implement a polarization-conserving channel.

\section{Signal and Noise}

Two crucial points for any communication system are the amount of attenuation of the link and the noise introduced in the system. This is even more important for quantum communication since the signal transmitted by Alice is ideally one photon (or a weak coherent pulse with very low mean photon number in many realistic implementations). Therefore one cannot increase the signal power in order to have a good enough SNR: the only available tools are the reduction of the link attenuation and of the background noise. In this section we will analyze a quantum channel between a ground station and a LEO satellite both in the uplink and the downlink, presenting a model for the expected attenuation and background noise.

\subsection {Signal attenuation}

The main factor limiting the performance of free-space optical communication is atmospheric turbulence, both for terrestrial horizontal links or for links between ground and satellites. Atmospheric turbulence induces refractive index inhomogeneities, that increase the amount of spreading for traveling beams \cite {Fante1980}. In particular, turbulent eddies whose size is large compared to the size of the beam induce a deflection of the beam (beam wandering), while smaller-scale turbulent features induce beam broadening. In other words, observing a beam which propagates through turbulent atmosphere at different time instants, one can see a broadened beam randomly deflected in different directions. When integrating the observation over a time-scale longer than the beam-wandering characteristic time, the global effect is a large broadening of the beam.

Models for optical beam propagation in the case of uplinks and downlinks between a satellite and a ground station have been discussed in the literature \cite {andrewsAO95, diosAO04, toyodaAO05}. In the case of a Gaussian beam of waist $w_0$ and intensity $I_0$, the average long-term spot (which is a superposition of moving short-term spots), tends theoretically to a Gaussian spatial distribution of intensity \cite {diosAO04}:
\begin {equation}
\left \langle I(r, L) \right \rangle = I_0 e^{-2r^2/w^2_{LT}}
\end{equation}
with width $w_{LT}$, where
\begin {equation}
w^2_{LT} = w^2_{ST} + 2 \left \langle \beta^2 \right \rangle
\end {equation}
Here $w_{ST}$ is the short-term beam width, while $\beta$ is the instantaneous beam displacement from the unperturbed position.

It can be shown that, for a collimated beam, the long-term beam width is \cite {diosAO04}:
\begin {equation}
w^2_{LT} = w^2_0 \left( 1 + \frac{L^2}{Z_0^2} \right) + 2 \left( \frac{4L}{k r_0} \right)^2
\end {equation}
where $Z_0$ is the Rayleigh parameter of the beam, $L$ is the propagation distance and $r_0$ is the Fried parameter, given by:
\begin {equation}
r_0 = \left [  0.42k^2 \int_0^L C_n^2 (z) \left( \frac{L-z}{L} \right)^{5/3} dz \right]^{-3/5}
\end {equation}
The estimate of r0 in equation (4) was made by integrating the turbulent contribution of the atmosphere along the whole optical path. The resulting $w_{LT}$ should then be considered a high bound, and the resulting conclusions as on the safe side. The refractive index structure constant $C_n^2(z)$ is taken from Ref. \cite{andrewsAO95} to be:
\begin{equation}
C_{n}^{2}(h)=0.00594(v/27)^{2}(h\cdot10^{-5})^{10}\cdot e^{-h/1000}+2.7\cdot10^{-16}e^{-h/1500}+A\cdot e^{-h/100}
\end{equation}
where $A=1.7\cdot10^{-14}$ and $v=21m/s$.
The expression for the short-term width is:
\begin{equation}
w^2_{ST} = w^2_0 \left( 1 + \frac{L^2}{Z_0^2} \right) + 2 \left \{ \frac{4.2L}{k r_0} \left[ 1 - 0.26 \left( \frac{r_0}{w_0} \right)^{1/3} \right] \right \}^2
\end{equation}

The receiving telescope can be described as a circular aperture of radius $R$, which collects part of the incoming beam and focuses it on a bucket single-photon detector. The power $P$ received through a circular aperture of radius $R$ centered on the beam is:
\begin {equation}
P = 2 \pi I_0 \int_0^R \rho e^{-2 \frac{\rho^2}{w_{LT}^2}} d\rho
\end {equation}

\begin{figure}[htbp]
\begin{center}
\includegraphics[width = 15 cm]{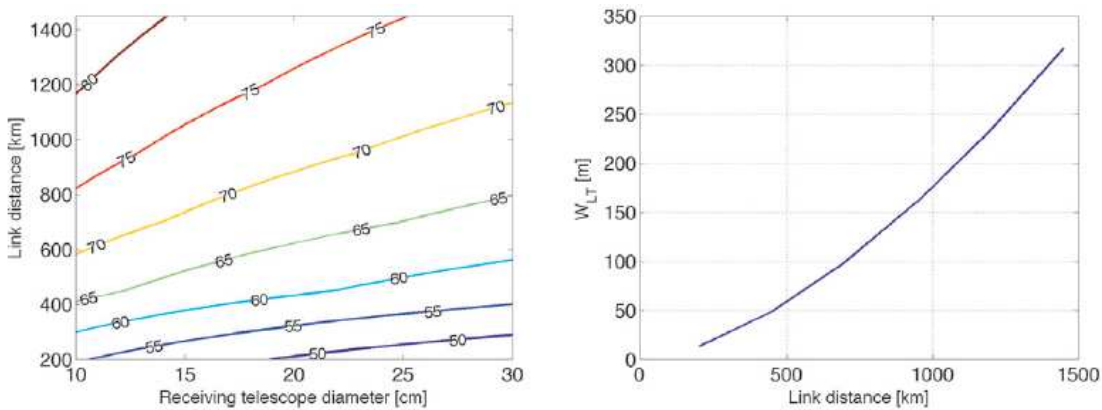}
\caption{\footnotesize Attenuation $\eta$ of the link (dB) as a function of the link distance and receiver telescope diameter for the long-term beam spreading effect, which takes into account the effects of beam spreading and wandering. The operating wavelength is $\lambda=800 nm$ and the diameter of the Earth-based transmitting telescope is assumed to be 1.5 m. On the right-side a zoom of the figure on the left for the link distance between 200 and 500 km.}
\label{figure1}
\end{center}
\end{figure}

Therefore the link-efficiency $\eta$, which we define as the percentage of the received power with respect to the transmitted one is:
\begin{equation}
\eta = \eta_0 \left( 1 - e^{-\frac{2R^2}{w^2_{LT}}} \right)
\end{equation}

The factor $\eta_0$ comprises the detection efficiency, the pointing losses and the atmospheric attenuation; we take an empirical factor \cite{aspelmeyerIEEE03} $\eta_0 \approx 0.1$.

Some simulations for the link efficiency are shown in Fig. \ref{figure1}: the link attenuation (in decibels) is shown as a function of the link distance $L$ and the radius $R$ of the receiving telescope. In the uplink the beam first travels through the turbulent atmosphere and then propagates, aberrated, in the vacuum to the satellite. The initial atmosphere-induced aberrations greatly increase the beam spreading, resulting in a very strong attenuation. For a relatively low satellite, at 500 km above the Earth surface, the attenuation is on the order of more than $50$ dB.

\begin{figure}[htbp]
\begin{center}
\includegraphics[width = 15 cm]{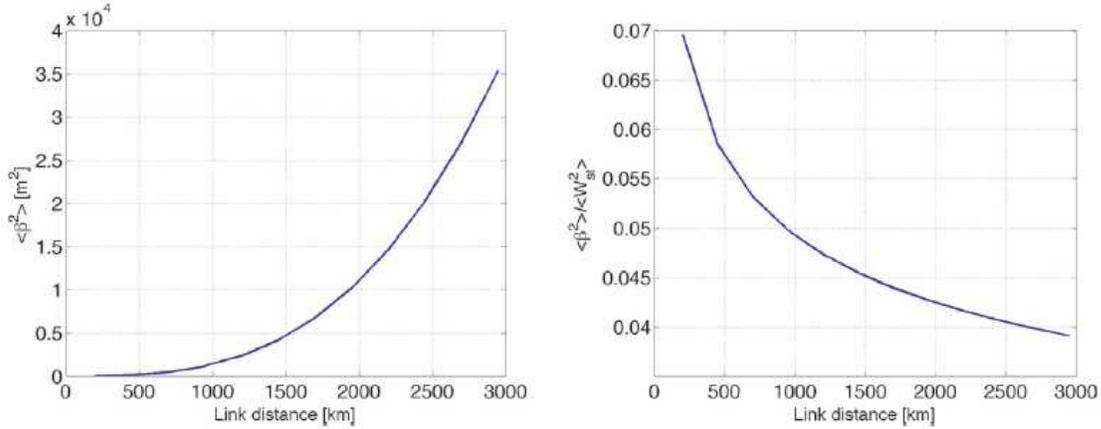}
\caption{\footnotesize On the left, $\left \langle \beta^2 \right \rangle$ as a function of the ground-to-satellite distance. On the right, ratio between $\left \langle \beta^2 \right \rangle$ and $\left \langle w^2_{ST} \right \rangle$. In the case of a LEO satellite, the effect of beam wandering is clearly limited to less than $10$ percent with respect to the beam spreading; therefore its possible compensation with a tip/tilt active system might not significantly improve overall performance of the link.}
\label {figure2}
\end{center}
\end{figure}

An interesting point is the relative contribution of the beam spreading due to smaller-scale atmospheric turbulence (described by $w_{ST}$) and the the beam-wandering induced by larger-scale eddies (described by $\left \langle \beta^2 \right \rangle$). In principle the beam wandering could be compensated by means of an active tip/tilt mirror with some kind of feedback loop. However, as it is shown in Fig. \ref{figure2}, the benefit that one could get is below $10$ percent, making such compensation practically worthless.

\subsection{Noise}

\subsubsection {Up-link (day-time operation)}

\begin{figure}[htbp]
\begin{center}
\includegraphics[width = 11 cm]{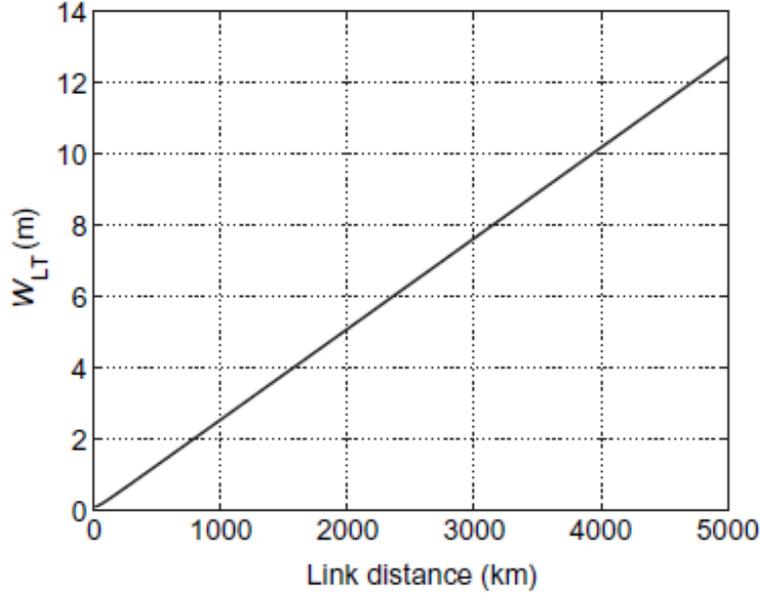}
\caption{\footnotesize Scheme to calculate the background noise in the uplink. Sun (or Moon) light is reflected by the Earth surface (with Lambertian diffusion) into the receiving telescope field-of-view.}
\label{figure3}
\end{center}
\end{figure}

During the day the main source of background noise is the sunlight reflected by the Earth surface into the telescope field of view (see Fig. 3).
Let $H_{sun}$ be the solar spectral irradiance (photons/ $s$ $nm$ $m^2$) at one astronomical unit and $a_E$ the Earth albedo; assuming a Lambertian diffusion, for which the radiance is independent of the angle, the spectral radiant intensity reflected by the Earth (number of photons per $s$, $nm$ and $sr$) is:
\begin{equation}
S_E = \frac{1}{\pi} a H_{sun} \Sigma
\end{equation}
where $\Sigma$ is the emitting area seen by the telescope and $H_{sun} = 4.61 \cdot 10^{18}$ photons/ ($s$ $nm$ $m^2$) at $\lambda = 900$ nm.
Such photons are collected by an optical system having entrance aperture raius $r$ and instantaneous field-of-view $IFOV$, at distance $L$ from the Earth surface.
Therefore the emitting area is:
\begin{equation}
 \Sigma = (\mbox{IFOV})^2 L^2
\end{equation}
and the solid angle from which the telescope on the satellite can be seen from Earth is:
\begin{equation}
\Omega = \frac{\pi r^2}{L^2}
\end{equation}
Therefore the number of background photons collected by the optical system per units of $\Delta \nu$ and $\Delta t$ is:
\begin {equation}
N_{day} = \frac{1}{\pi} a_E H_{sun} \Sigma \Omega = a_E r^2 (IFOV)^2 H_{sun}
\end {equation}

\subsubsection{Uplink (night-time operation)}
The dominant sources of background radiation from the Earth surface during night are its black-body emission and the reflected moon light. In realistic situations there will be a significant contribution of scattered light from human activities, which depends on the specific location considered.\\
The number of photons per ($s$ $nm$ $m^2$) emitted by a black body, according to Planck's law, is:
\begin{equation}
 S_{bb} = \frac{2c}{\lambda^4} \frac{1}{e^{\frac{hc}{\lambda k T}}-1}
\end{equation}
At $T = 293$ K and $\lambda = 800$ nm, $S_{bb} = 3.1 \cdot 10^6$ photons per ($s$ $nm$ $m^2$).

Let's now calculate the radiance due to moonlight reflection on the Earth. Given the solar spectral irradiance $H_{sun}$, the number of photons per $s$ and $nm$ which hit the Moon surface is: $H_{sun} \cdot \pi R^2_M$ where $R_M$ is the Moon radius. Assuming Lambertian diffusion, the number of photons per ($s$ $nm$ $sr$) reflected by the Moon is:
\begin{equation}
\tilde{P}_{moon} = \frac{a_M}{\pi} S_{sun} \pi R^2_M
\end{equation}
where $a_M$ is the Moon albedo.
Assuming the Moon at normal incidence, the solid angle to the area on Earth $\Sigma$ seen by the telescope is:
\begin{equation}
\Omega_{\Sigma} = \frac{\Sigma}{d^2_{EM}}
\end{equation}
where $d_{EM}$ is the distance Earth-Moon.
The spectral radiant intensity after Lambertian reflection from the Earth surface is:
\begin {equation}
S_E^{(M)} =  \frac{1}{\pi} a_E a_M S R_M^2 \frac{\Sigma}{d^2_{EM}}
\end {equation}
The number of photons per second and $nm$ of bandwidth entering the receiving telescope (radius $r$, field-of-view $IFOV$) is:
\begin{equation}
N_{night} = \alpha N_{day}
\end{equation}
where:
\begin {equation}
\alpha = a_M \left( \frac{R_M}{d_{EM}} \right)^2
\end{equation}
is the ratio between the background radiance at night-time (full Moon) and day-time. Assuming the Moon albedo to be $a_M \approx 0.12$ we have that $\alpha$ is of the order of $10^{-6}$: during night-time, in full Moon conditions, we have approximately a reduction of six orders of magnitude in the amount of background noise.

\begin{figure}[htbp]
\begin{center}
\includegraphics[width = 11 cm]{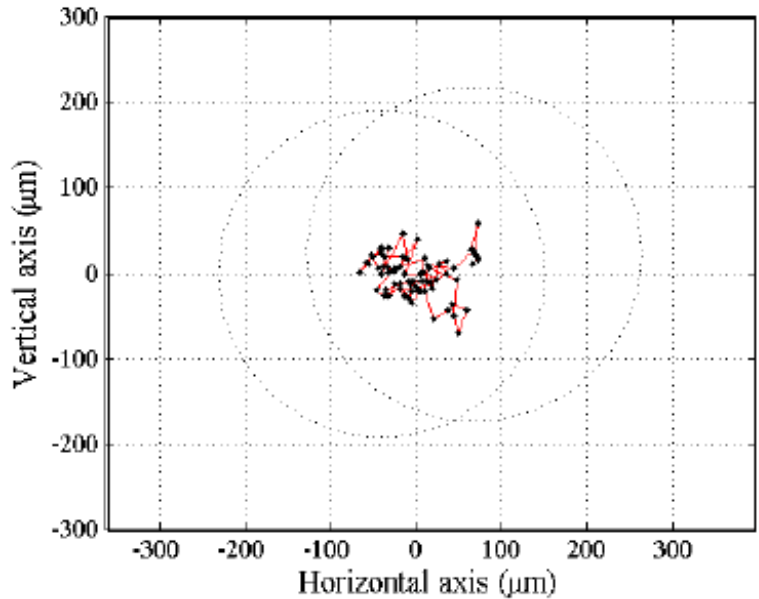}
\caption{\footnotesize Number of photons at day-time (left side) and night-time (right side) as function of telescope IFOV where $\Delta\nu=1 nm$, $\Delta t=1 ns$. The number of background photons entering the telescope during night-time in full-Moon conditions is approximately six order magnitude larger than the value for day-time operation.}
\label{figure4}
\end{center}
\end{figure}

\begin{figure}[htbp]
\begin{center}
\includegraphics[width = 11 cm]{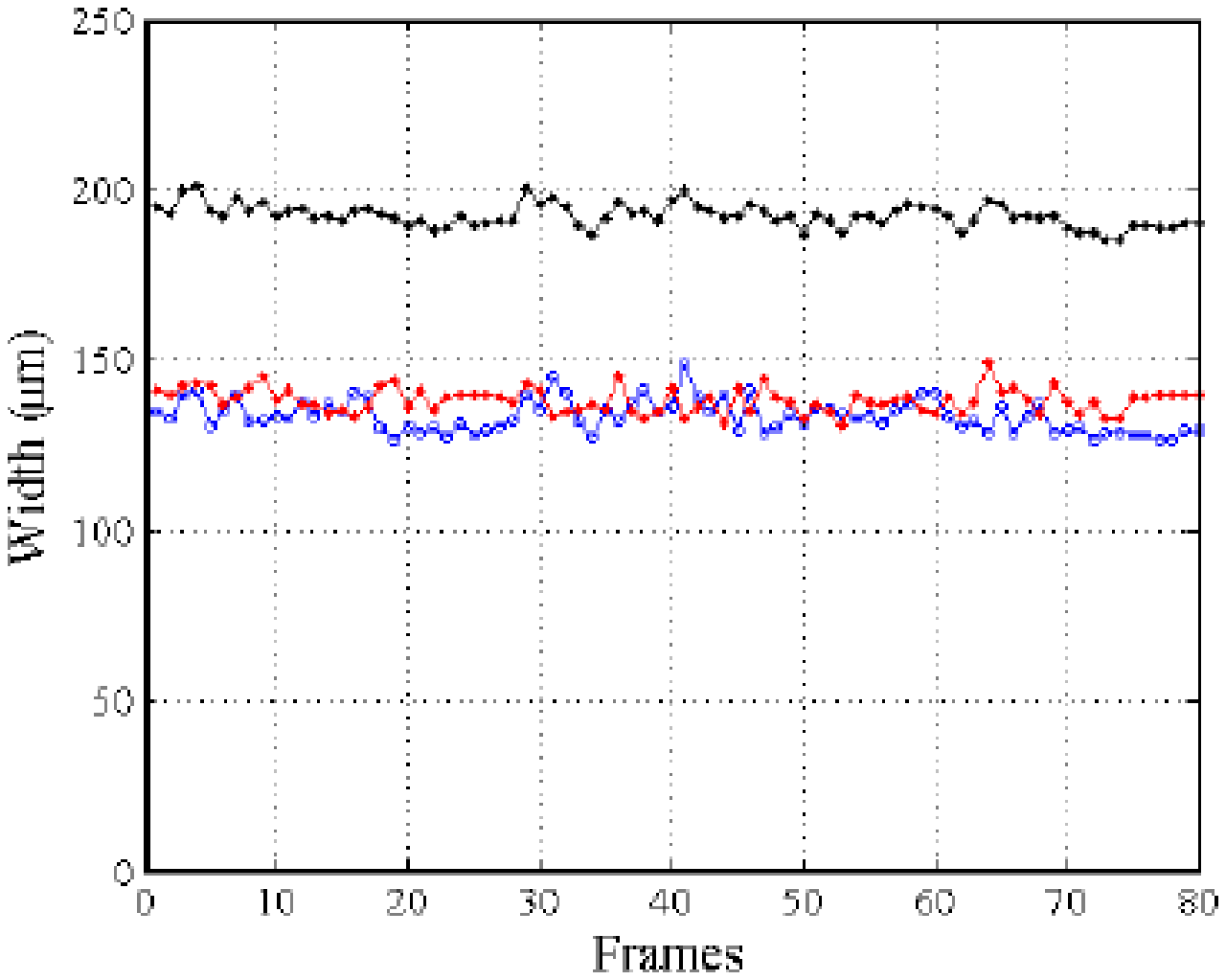}
\caption{\footnotesize Signal-to-noise ratio (in dB) at day-time (left side) and night-time (right side) as a function of telescope IFOV and satellite distance. The curves on the left sign correspond to negative values for the SNR in decibels; this means that the SNR is lower than 1, clearly to low to establish a quantum communication link. On the other hand, SNR as high as 100:1 or 1000:1 can be envisaged during night-time. The operating wavelength is $\lambda=800$ nm and the transmitting telescope diameter is $1.5$ m. We assume a filtering bandwidth $\Delta\nu=1$ nm and a gating time of $\Delta t=1$ ns for the detectors.}
\label{figure5}
\end{center}
\end{figure}

\subsection{Down-link}

\subsubsection {Signal attenuation and turbulence}

The effect of the atmospheric turbulence on a plane wavefront is a phenomenon very relevant to the FOV limit in the noise reduction of a quantum channel, as seen above. In particular, the predominance of broadening of the beam or of the rapid bending of the beam, described by the long- and short-terms in the far-field width is a crucial information in order to design the optical system aimed at the mitigation of the turbulence effects.

To this purpose, experimental data taken by means of a ground telescope are suitable to be compared to the modeling presented before. In our experiment we have acquired with a video recorder the flickering light from Vega ($\alpha$-Lyrae, magnitude=0) by the Matera Laser Ranging Observatory of Agenzia Spaziale Italiana in Matera, Italy. The telescope has the primary mirror diameter of 1.5 m. The gathered light was spectrally filtered in the
green by the coated optical components of the Coudè  path, and acquired on the focal plane by a bidimensioanl sensor whose square pixel size was of 6.7 $\mu$m.  The collection of the frames were analyzed in order to extrac the first two moments of the intensity distribution. The results is reported in the Figs .. ..

\subsubsection{Background light noise}
The background noise for a satellite-to-ground link was examined in details by Miao et al. \cite{miaoNJP05}. The noise power $P_{b}$ received by a ground-based telescope pointing a satellite in the sky can be expressed as:

\begin{equation}\label{B1}
	P_{b}=H_{b}\cdot\Omega_{fov}\cdot \pi r^2 \cdot \Delta \nu
\end{equation}
where $H_{b}$ is the brightness of the sky background in $W\ m^{-2}\ sr^{-1}\ \mu m^{-1}$, $\Omega_{fov}$ the field of view of the telescope in $sr$ and $r$ its radius; $\Delta \nu$ is the filter bandwidth. $H_b$ is strongly related to the weather conditions, for example during.

\begin{figure}[htbp]
\begin{center}
\includegraphics[width = 11 cm]{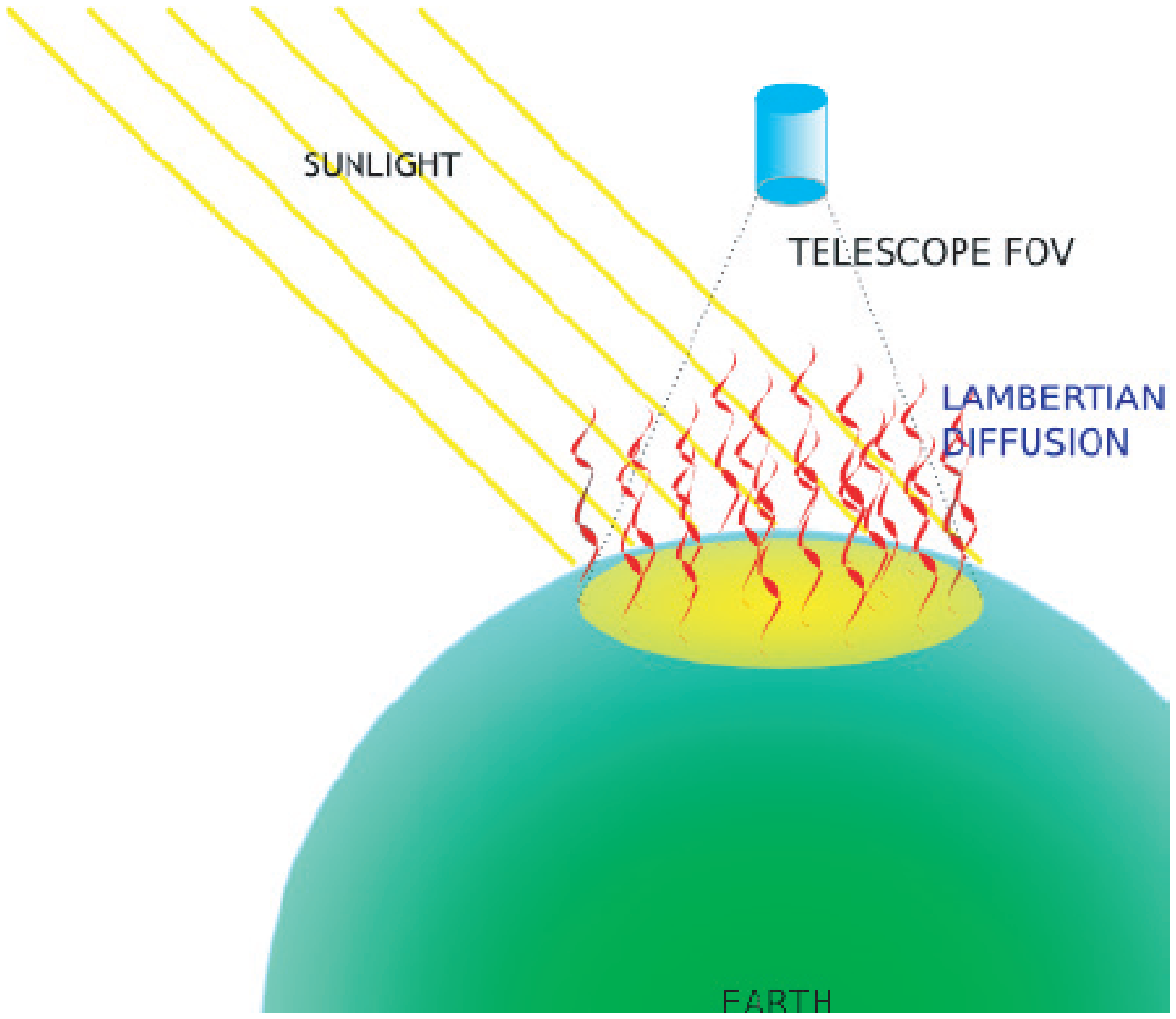}
\caption{\footnotesize Simulations for the downlink. On the left side, link attenuation (in dB), as a function of the diameter of the satellite-based transmitting telescope and of the link distance L. On the right side, SNR of the link as a function of the sky background noise for different link distances, assuming a diameter of the transmitting telescope $r=15$ cm. All  simulations are performed assuming a diameter of the Earth-based receiving telescope $R_X=1.5$ m, with field-of-view 0.016 degrees (corresponding to MLRO, Matera)}
\label{figure6}
\end{center}
\end{figure}

We calculated the signal-to-noise ratio for the downlink using our results for the signal attenuation in a turbulent atmosphere and the noise parameters given in \cite{miaoNJP05}. The results are shown in Fig. \ref{figure6}. On the left side, the down-link attenuation is shown as a function of the link distance $L$ and the radius of the Earth-based receiving telescope. Two factors result in an increased performance for the downlink with respect to the uplink. First, on Earth we can have larger receiving telescopes than in space. Second, the beam first propagates in the vacuum with just diffraction spreading and gets in contact with the turbulent atmosphere only in the final stage of propagation. Therefore the aberrations introduced by turbulence only affect weakly the wavefront before it enters the telescope.

On the right side of Fig. \ref{figure6} we plotted the SNR as a function of the sky brightness ($\Delta \nu = 1$ nm, $IFOV$). The SNR is greater than one only at night-time.

\subsection {Filtering and synchronization}
As discussed in the previous sections, in order to have a significant signal-to-noise ratio to establish a quantum communication link, it is crucial to reduce the amount of noise. Moreover, the management of detector dead-time is also a crucial point. Avalanche single-photon detectors are characterized by a certain amount of time, after a detection event, in which they cannot detect any more photons. For single-photon avalanche photodiodes the dead time can vary from 40 to 100 ns. For this reason the detection of a noise photon has a double negative effect: it decreases the final secure key rate and it blinds the detector for the duration of the dead time preventing the detection of a good photon. This is why very often QKD systems run in gated-mode: the detectors are switched-on only when a signal photon is supposed to arrive. To allow the possibility to gate the detectors, high-accuracy temporal synchronization is mandatory.

Filtering strategies can be divided in three categories: spectral, spatial and temporal. Spectral filtering is pretty easy to implement even on a Space setting, just employing interference filters which must be thermally stabilized. Spatial filtering can be implemented acting on the receiving telescope field-of-view, in order to select only photons coming from the right directions. A trade-off must be found between the need of a strong spatial selectivity (to have efficient noise-reduction) and the possibility of imperfect pointing, which would call for a relaxation in spatial filtering.

\begin{figure}[htbp]
\begin{center}
\includegraphics[width = 15 cm]{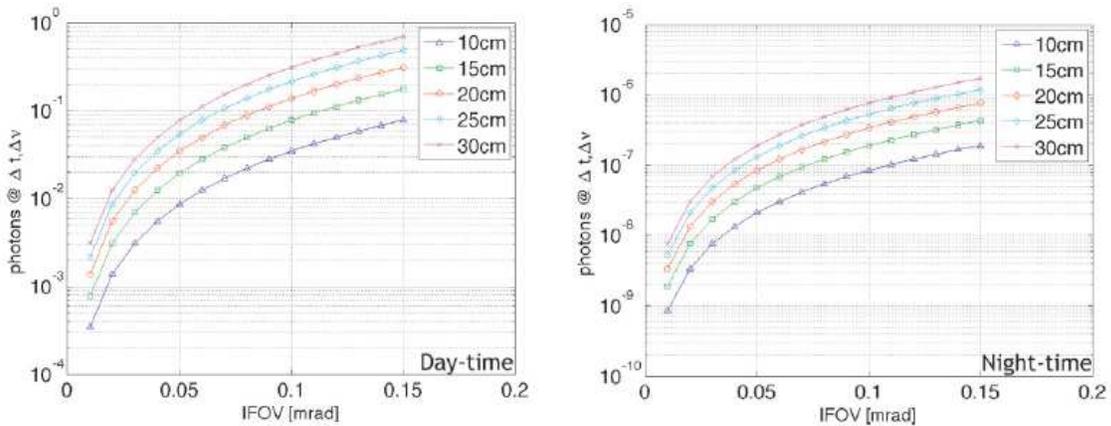}
\caption{\footnotesize Data analysis of the KHz laser-ranging data of a typical passage of the GRACE satellite (altitude around 400 km). On the left side time derivative of the range time as a function of time, on the right-side histogram of the same quantity. The temporal variation of the range time is around $40 \mu s/s$.}
\label{figure7}
\end{center}
\end{figure}

In a Space environment, time filtering is a more delicate issue, because it requires really good synchronization between two devices in fast relative motion. Such precise synchronization is fundamental in order to discriminate the good photons from the background noise and in order to have gated-mode operation of the detectors. For the latter, the arrival time of a signal photon need to be known in advance, in order to open the detector gate. Let's focus to the case of an Earth-based station sending photons to an orbiting receiver. Two different schemes have been used in the literature to synchronize free-space QKD systems: self-synchronization and external synchronization.

In the case of self-synchronization a periodic bright pulse of a wavelength different from the one of the signal photons can be used to open the detector gate. This technique was used, for example, in one of the seminal experiments about free-space QKD by Hughes and co-workers \cite{hughes02}. The waveform of the pulses can be shaped in order to code in the synchronization frames some information regarding the communication itself.
A different option is to use an external synchronization technique, for example stabilized local clocks and software-controlled phase-lock loop driven by the detected photon signal (as described by Rarity et al. in \cite{rarityJMO01})  or by the global positioning system (GPS) signal (as done R. Ursin and coworkers for entanglement distribution over 144 Km in free-space \cite{ursin07}). In the case of satellite-based quantum communication this technique requires the precise a priori knowledge of the orbit, which makes it extremely unpractical.

Here we will focus on the self-synchronization technique, which we believe is easier to realize in practice and gives the signal for the detector-gating control with no need to know precisely the station-satellite distance. An interesting question is then what repetition rate shall be imparted to the synchronization pulses in order to keep a control on the satellite position on the order of the tens of centimeters (which correspond to trip-times to the order of one nanosecond).
In Fig. \ref{figure7} we show some data analysis performed on KHz laser ranging data for the GRACE satellite acquired by the Graz Space Research Center. From the laser-ranging data we calculated the time derivative of the photon trip-time from ground to the satellite and we plotted it on a histogram. The results clearly show that the trip-time changes of the order of $40 \mu s$ per second (which corresponds to about 12 Km per second). This means that if we want to keep track of the changes with an accuracy to the order of the nanosecond we need a repetition rate for the synchronization of the order of $50-100$ KHz.

\section {Polarization Control}

A second crucial point for the implementation of quantum communication schemes based on polarization-encoded qubits is, of course, the preservation of polarization states in the channel.

\begin{figure}[htbp]
\begin{center}
\includegraphics[width = 15 cm]{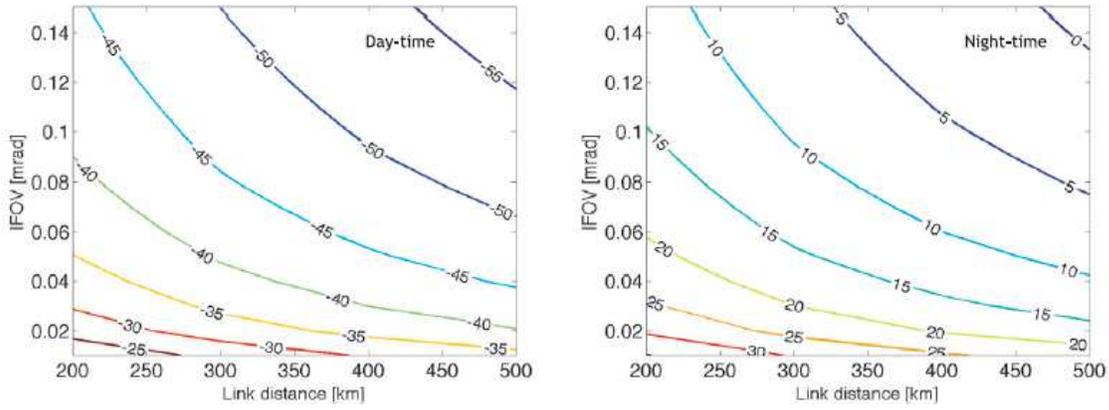}
\caption{\footnotesize Scheme of the satellite tracking system and its effect on polarization, as discussed in \cite{bonatoOE06}. A source on a satellite emits a stream of single photons, which are directed to ground by a moving pointing mirror. A second pointing mirror on ground receives the photons and whatever direction they come from, it sends them to the detection apparatus. Due to the relative motion between the satellite and the ground station, there is a relative rotation of the polarization axes and a change in the mirror incidence angles, which induces a time-dependent polarization transformation on the qubits.}
\label{figure8}
\end{center}
\end{figure}

\begin{figure}[htbp]
\begin{center}
\includegraphics[width = 15 cm]{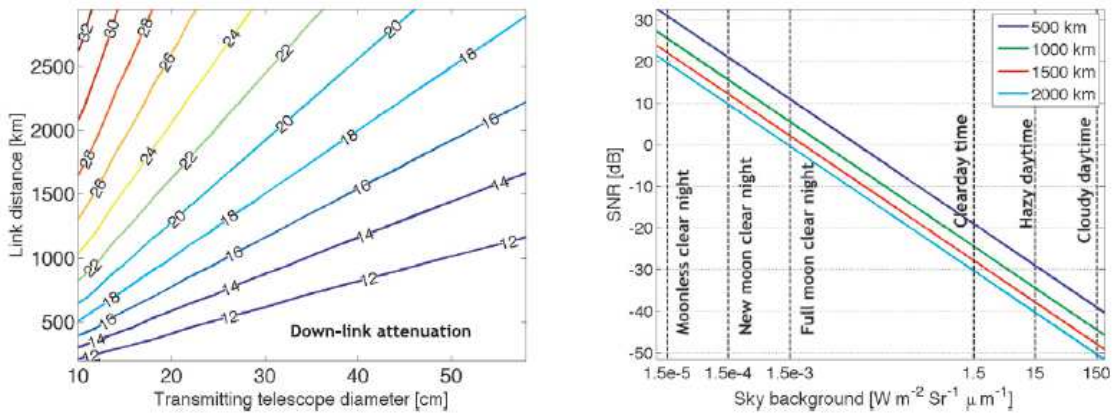}
\caption{\footnotesize Example of temporal evolution of the Stokes parameters at the receiver for a fixed vertically-polarized state emitted on the satellite. In a few-minutes passage the values of the Stokes parameters change dramatically in a smooth way.}
\label{figure9}
\end{center}
\end{figure}

As it was shown in \cite{bonatoOE06}, propagation in the atmosphere does not affect significantly the polarization stets, nor does the Faraday effect due to the Earth magnetic field. The use of curved optics in an off-axis configuration introduces some spatially-dependent polarization effects \cite{bonatoIOP07} which can lead to global decoherence of the polarization-encoded qubits. However, the effect is small for on-axis optics and it can be neglected, just having some care in the design of optical systems.

On the other hand, the relative motion of the satellite and the ground station induces a time-dependent transformation on the polarization state as seen by the receiver. This is mainly due to the relative rotation of the satellite vertical axis with respect to ground and change in the polarization induced by reflection on mirrors at time-varying angles. The effect, in the case of a single passage of a LEO satellite orbiting at 400 Km from the Earth surface, is shown in Fig. \ref{figure9}: given a photon which is emitted with vertical-polarization in the satellite reference frame, the Stokes parameter seen by the ground-based receiver are plotted as a function of time.

If we can neglect channel depolarization effects, as it is the case for atmospheric propagation, polarization states can be represented by Jones vectors:
\begin {equation}
\left[
\begin{array}{c}
A \\
B e^{i \varphi}
\end{array}
\right ]  \qquad A, B, \varphi \in R \quad A^2+B^2=1
\end{equation}
The channel properties are described by a 2-by-2 time-dependent Jones matrix $C (t)$, which transforms the polarization states according to $J(t) = C(t) J_0$. To establish a successful quantum link based on polarization-encoded qubits, such transformation must be compensated. This can be done characterizing the channel without interfering with the single-photon exchange, in order to measure such matrix $C (t)$. Then, applying the inverse transformation $C^{-1} (t)$ for every time instant $t$ to the incoming photons, the correct state can be restored before performing the measurements needed fro quantum key distribution.

However, in general, not to interfere with the signal photon exchange, the characterization of the channel Jones matrix is to be performed with different parameters than the photon exchange. For example, a different wavelength may be employed, or the two operations of channel-probing and quantum-communication can be performed at in different time-slots. Defining $C_P (t)$ as the experimentally measured channel Jones matrix, we have:
\begin{equation}
C_P^{-1} (t) C(t) = E(t)
\end{equation}
with $E(t) \rightarrow I, \quad \forall t$ in the case of ideal compensation. Let $\{E_{ij} (t) \}, i,j = 1,2$ be the elements of the matrix $E(t)$.

In this Section we will discuss some polarization-compensation schemes, discussing their effectiveness in the case of the model presented in \cite{bonatoOE06}. Considering a BB84 quantum key distribution scheme, photons are transmitted in two non-orthogonal bases, for example the horizontal/vertical one (states $\left | H \right \rangle$ and $\left | V \right \rangle$) or the diagonal one (states linearly polarized at $\pm 45$ degrees, that we will indicate respectively with $\left | + \right \rangle$ and $\left | - \right \rangle$ degrees).
The average error probability is:
\begin {equation}
P_E = 1 - P_{HH} - P_{VV} - P_{++} - P_{--}
\end {equation}
where $P_{ij}$ is the temporal average of the conditional probability of measuring the state $i$ once $j$ has been transmitted ($P_{ij} = \left \langle p (r = i | t = j)\right \rangle$).

Suppose now to have a horizontally-polarized state transmitted at time $t_i$. After compensation, one gets the state:
\begin {equation}
J (t_i) =
\left [
\begin{array}{cc}
E_{11}(t_i) & E_{12} (t_i) \\
E_{21} (t_i) & E_{22} (t_i)
\end{array}
\right ]
\left [
\begin{array}{c}
1 \\
0
\end{array}
\right ]
=
\left [
\begin{array}{c}
E_{11} (t_i)\\
E_{21} (t_i)
\end{array}
\right ]
\end{equation}
so that the probability of obtaining the correct result is $|E_{11} (t_i)|^2$. Assuming that the probability of transmitting a $\left | H \right \rangle$ state is $\frac{1}{4}$:
\begin {equation}
P_{HH}= \frac{1}{4} \left \langle |E_{11}|^2 \right \rangle
\end {equation}
With similar arguments one can find expressions for the other conditional probabilities so that:
\begin {equation}
P_E = 1 - \frac{1}{8} \left \langle \left \{ 3|E_{11}|^2 + 3|E_{22}|^2 + |E_{21}|^2 + |E_{12}|^2 + E^*_{11}E_{22} + E_{11}E^*_{22} + E^*_{12}E_{21} + E_{12}E^*_{21} \right \} \right \rangle
\end{equation}

\subsection {Probe beam at a different wavelength}
One possible way of measuring the channel Jones matrix without perturbing the single-photon exchange is using a probe beam at a wavelength $\lambda_p$ different from the one of the signal beam ($\lambda_s$). In this case the signal transformation Jones matrix is $C(\lambda_s)$, while the compensation matrix is $C(\lambda_p)$. Thefore:
\begin {equation}
E = C^{-1} (\lambda_p) C (\lambda_s)
\end{equation}

\begin{figure}[htbp]
\begin{center}
\includegraphics[width = 15 cm]{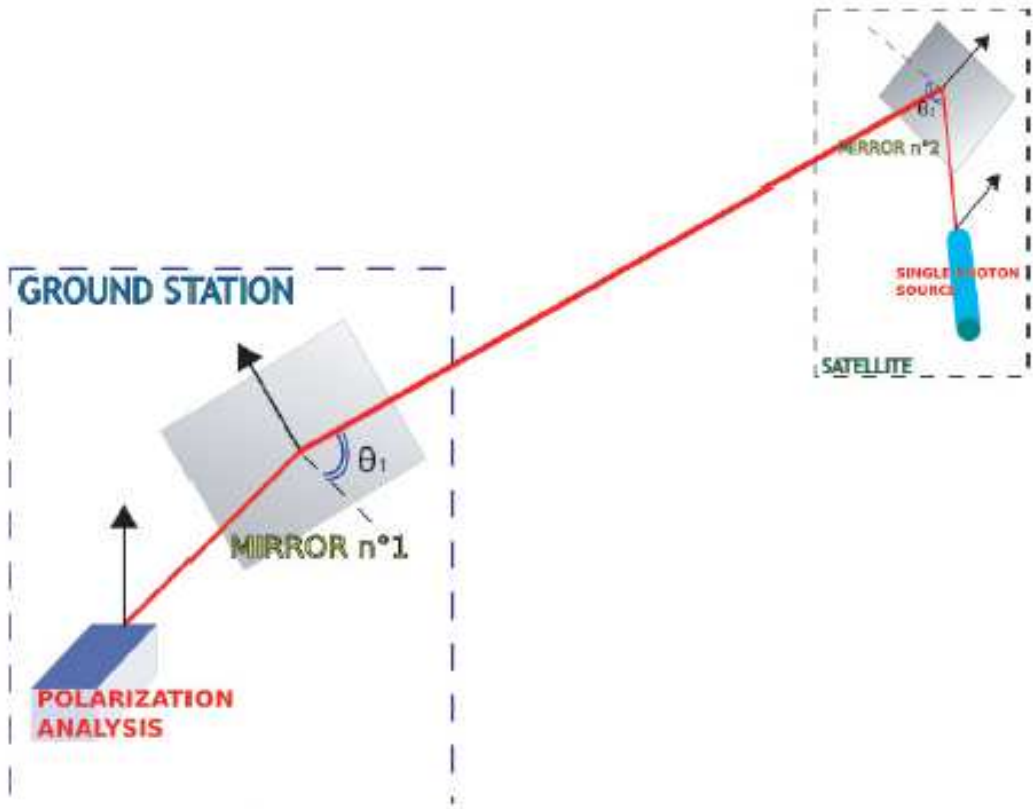}
\caption{\footnotesize Polarization-state preservation in the satellite-based quantum channel, in the case of a channel-probing beam at a different wavelength with respect to the signal. The bit error rate due to imperfect compensation is negligible even for probing wavelengths quite far from the signal one ($\lambda_s = 800$ nm). }
\label{figure10}
\end{center}
\end{figure}

To have a statistical evaluation of the degree of compensation that can be achieved with this technique we performed a simulation for 1000 passages of a LEO satellite orbiting at 500 Km. We used the model described in \cite{bonatoOE06} to calculate the matrices $C(\lambda_s)$ and $C (\lambda_p)$ for a uniform temporal sampling of each passage ($\Delta t = 1$ s). Then we computed for each time instant the matrix $E$ and the error probability $P_E$, finally averaging over time. The results are reported in Fig. \ref{figure10}, showing the QBER due to compensation error as a function of $\lambda_P$. Perfect compensation is clearly possible using a wavelength for the probe beam very close to that of the signal beam. However an acceptable error rate (below one percent), is possible for wavelengths much more distant than our signal one.

\subsection {Time-multiplexing of signal and probe beam}
A different compensation scheme can be time-multiplexing of signal and probe pulses at the same wavelength in the channel. In this case, suppose to send the probe pulses with repetition rate $f_{P}$, so that the m-th probe pulse will be emitted at time $t_m^{(0)} = m T_{0}$ with $T_0 = 1/f_{P}$. Between any two probe pulses, N single-photon pulses will be transmitted, each at the time $t_{m, i} = t_m^{(0)} + i \delta$ where $\delta = T_{0}/N$.
In other words, we measure the channel Jones matrix $C(t_m^{(0)})$ and use it to compensate N subsequent single-photon pulses:
\begin{equation}
J \left[ t_m^{(0)} + i \delta \right] = C^{-1}(t_m^{(0)}) J_0 \left[  t_m^{(0)} + i \delta \right]
\end{equation}

The repetition rate of such pulses must be fast enough to characterize in real-time the evolution of the channel properties. Assuming that this is the case, the amount of change for a Stokes parameter $S_j(t)$ at a time $\Delta t$ slightly after $t_m^{(0)}$ is small and can be expressed with a Taylor expansion to the first order:
\begin{equation}
S_j (t_m^{(0)} + \Delta t) - S_j (t_m^{(0)}) \approx \frac{dS_j}{dt} \vert_{t = t_m^{(0)}} \Delta t
\end{equation}
If we want to keep the error on $\Delta S_j(t)$ under a certain value $\Delta S_{max}$, the repetition rate of the probing pulses must be:
\begin{equation}
f_{P} \geq \frac{1}{\Delta S_{max}} \frac{dS}{dt} \vert_{max}
\end{equation}

\begin{figure}[htbp]
\begin{center}
\includegraphics[width = 11 cm]{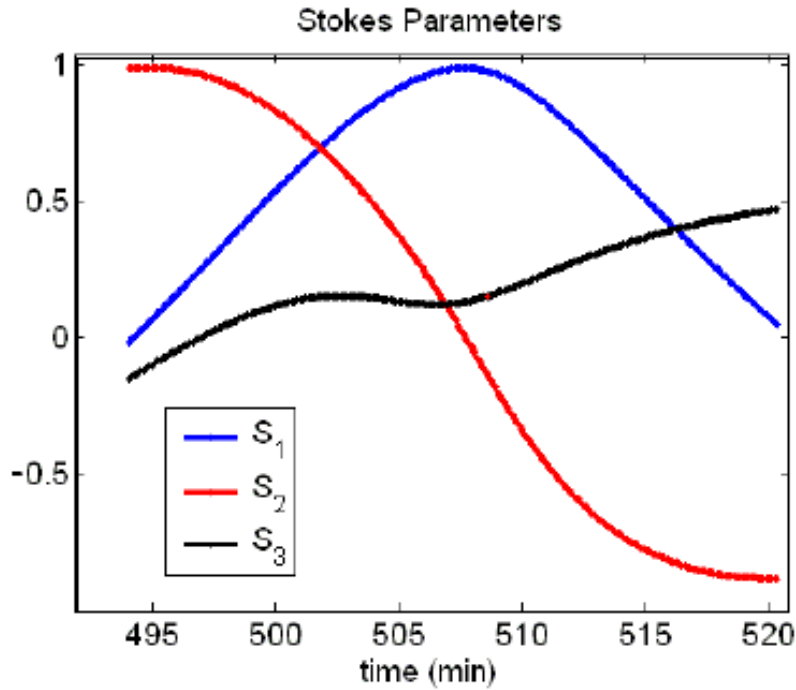}
\caption{\footnotesize Statistics of the time-derivative of the Stokes parameter $S_2$ for 1000 passages of two satellites with different orbits (500 Km for the picture on the left, 5000 Km for the one on the right). The temporal evolution of the transformation is faster for the lower-orbit satellite (the absolute value of the time derivative is within 0.015 $s^{-1}$). For higher satellite the transformation is slower (within 0.005 $s^{-1}$ for the figure on the right)}
\label{figure11}
\end{center}
\end{figure}

Assuming a maximum value for the time-derivative of the Stokes parameters of 0.02 (see Fig. \ref{figure11}), and stating for the maximum acceptable error on the Stokes parameters $\Delta S_{max} = 10^{-5}$, we get a value of $f_{P} = 2$ KHz for the probe repetition rate. This value is a large bound on the error, since $|dS/dt|$ is in general much smaller than the maximum value we took.

\begin{figure}[htbp]
\begin{center}
\includegraphics[width = 15 cm]{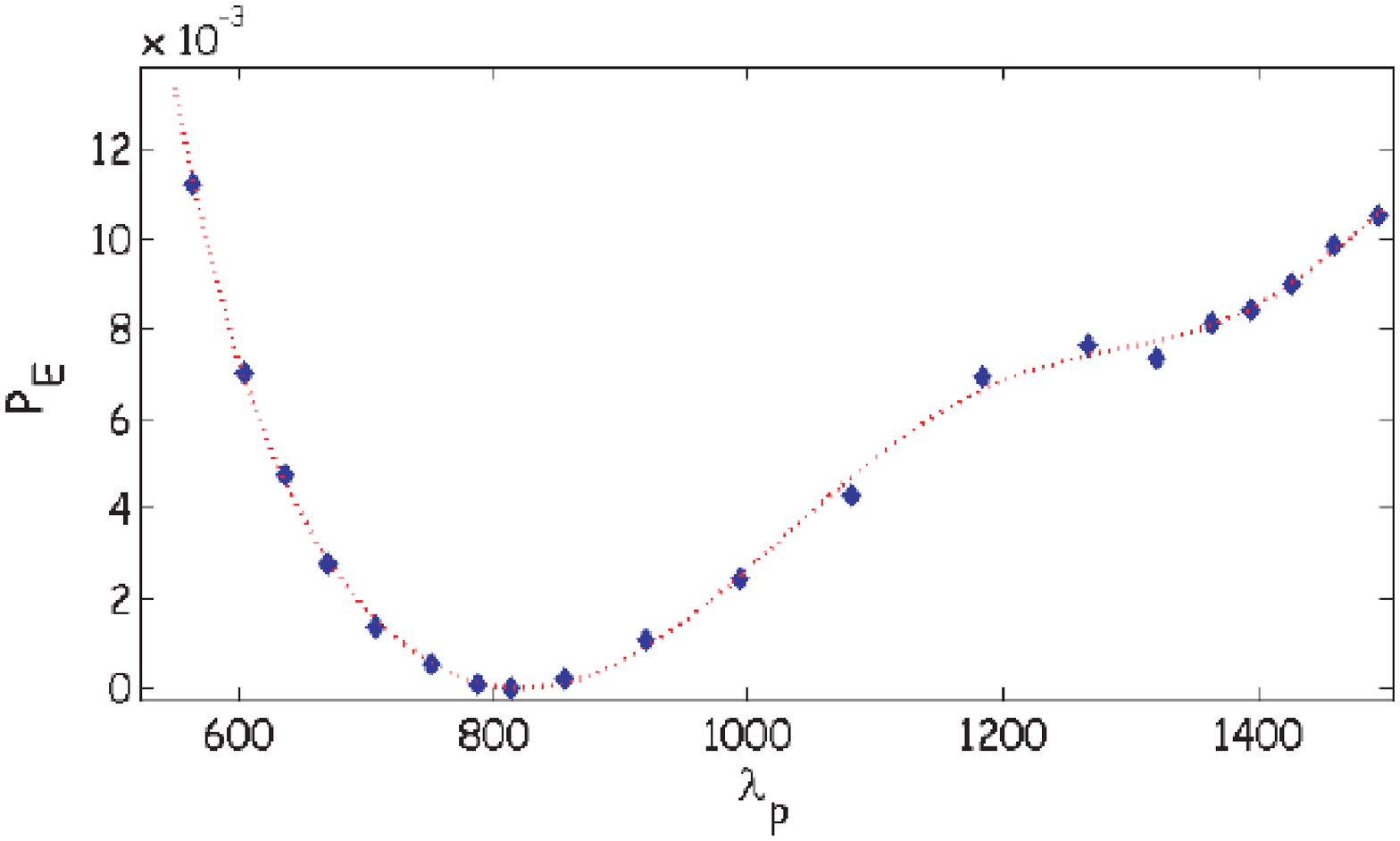}
\caption{\footnotesize Simulated quantum bit error rate due to imperfect polarization-compensation in the case of a temporal-multiplexing scheme, as a function of the channel-characterization pulses repetition rate. Data are shown for satellites at different altitudes.}
\label{figure12}
\end{center}
\end{figure}

To have a statistical evaluation of the average error probability we performed some simulations similar to what we did for the different-wavelengths scheme. In this case we computed the probability error as a function of the repetition rate of the probe pulses. The results are shown in Fig. \ref{figure12}.

\section {Discussion}
In this Section we will analyze the possibility of establishing a quantum key distribution link in different configurations employing a LEO satellite and an optical ground station, for different protocols.
Throughout the whole Section, formulas for key-generation rate in the asymptotic limit of a long key will be employed. This is clearly not true for real-world QKD experiments, especially in the cases involving the exchange of a secret key between a LEO satellite and Earth. At present, the analysis of the security of quantum key distribution for finite key lengths is still an open question [ref] in quantum information theory.

\subsection {BB84}
The secret key rate per pulse for the BB84 protocol in the case of an ideal single-photon source:
\begin{equation}
R_{BB84}^{(ideal)} \geq \frac{p_{exp}}{2} \left[ 1 - f(e)H_2(e) - H_2 \left(e \right)  \right]
\end{equation}
where $p_{exp}$ is the probability that a non-empty pulse is detected by Bob, $e$ is the QBER, $f(e)$ is the efficiency of error correction and $H_2(x)$ is the binary entropy function: $H_2(x) = -x \log_2 x - (1-x) \log_2(1-x)$. The efficiency of the classical error correction algorithm is described by the factor $f(e)$: we take $f(e) \approx 1.22$.

In most pratical quantum communication experiments, single photons are implemented with weak coherent pulses, for which there is a non-zero probability to produce multiphoton states. On such multiphoton pulses Eve could perform a photon-number-splitting (PNS) attack \cite{dusekOC99, BrassardPRL00, lutkPRA00}. She can split a photon from the multiphoton pulse, store it and measure it in the correct basis after Alice and Bob have pubblicly announced their bases. If she sends the rest of the multiphoton pulse to Bob no noise will be introduce in the channel and she can get complete information about the bit without being discovered. Such bits, that have leaked information to the eavesdropper, are called tagged bits. Inamory et al. \cite {ILM} and Gottesmann et al. \cite{GLLP} showed that in this situation a secure key can still be distilled and the key generation rate is given by:

\begin{equation}
R_{BB84} \geq \frac{p_{exp}}{2} \left[ (1-\Delta) - f(e)H_2(e) - (1-\Delta)H_2 \left(  \frac{e}{1-\Delta} \right)  \right]
\end{equation}

In the case of an uplink to a LEO satellite the channel is extremely lossy and almost all the single-photon pulses may be wasted, resulting in basically only multiphoton pulses giving clicks in Bob's detectors. Therefore, increasing the channel losses, the fraction of secure bits decreases. If the losses are so strong that only multiphoton pulses are detectd by Bob, no secure key can be generated.

As a worst-case estimate of the fraction of tagged bits $\Delta$ we can take the fraction of multiphoton pulses over the fraction on non-empty pulses detected by Bob \cite{BrassardPRL00}:
\begin{equation}
\Delta \approx \frac{1-e^{-\mu} - \mu e^{-\mu}}{1-e^{-\eta \mu}}
\end{equation}
In general, given a link attenuation $\eta$ the key generation rate is of the order of $O(\eta^2)$ (see \cite{WangPR07}).

\begin{figure}[htbp]
\begin{center}
\includegraphics[width = 15 cm]{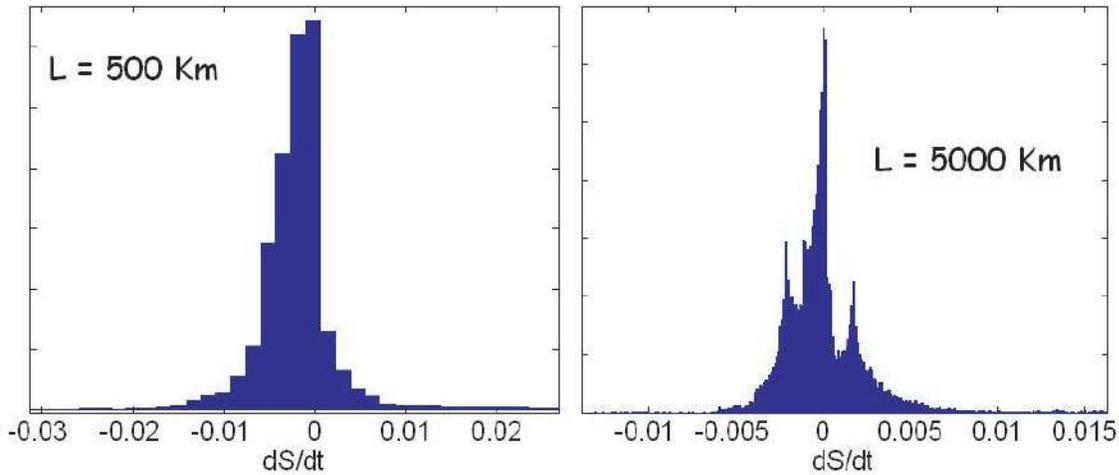}
\caption{\footnotesize Key-generation rate for the BB84 protocol using weak laser pulses as an approximate single-photon source. On the left side, results for the up-link, in the right side results for the downlink. For the uplink, the channel attenuation is so high that a QKD session with significant key generation rates cannot be implemented, while for the downlinka key-generation rate of $10^{-4}$ for a satellite orbiting at around 500 km can be obtained using a source with mean photon number $\mu = 0.01$. }
\label{figure13}
\end{center}
\end{figure}

Simulations for the key-generation rate as a function of the link distance are shown in Fig.  \ref{figure13}. In the case of the uplink the attenuation is so high that the secure key generation rate is extremely low (of the order of $10^{-12}$), on the other hand it is not possible to increase the value of $\mu$ in order to avoid PNS attacks.

For the downlink, on the contrary, a successful establishment of a BB84 QKD link is possible. Assuming $\mu = 0.01$ (see Fig. \ref{figure13}) and a source repetition rate of $10$ MHz, for a satellite at 500-600 km we can get around 1 kbit of secure key per second.

\subsection {Decoy-state}
To improve the performance of coherent-state weak-pulse QKD, the decoy state method has been proposed \cite{hwangPRL03, LoDecoyPRL05, wangDecoyPRL05}. For BB84 protocol, the security analysis is performed using a worst-case estimate on the fraction of bits that are known to the eavesdropper. The decoy-state technique, on the other hand, exploits states with different light intensities to probe the channel transmissivity and error probability, giving a more accurate bound on the amount of tagged bits.

Suppose to use a three-state decoy technique, which exploits vacuum states and two coherent states with mean photon number $\mu$ and $\mu'$. Let $S_{mu}$ be Bob's counting rate when Alice transmits pulses with mean photon number $\mu$ and $S_0$ be Bob's counting rate in the case of vacuum-state transmission (therefore due to dark counts and background noise). The bound for $\Delta$ is \cite{WangPR07}:

\begin{equation}
\Delta \leq \frac{\mu}{\mu'-\mu} \left( \frac{\mu e^{-\mu} S_{\mu'}}{\mu' e^{-\mu'}S_{\mu}} -1  \right) + \frac{\mu e^{-\mu} S_0}{\mu' S_{\mu}}
\end{equation}

Probing the channel with different light intensity we can get a more accurate estimate of $\Delta$. Consequently, we can guarantee unconditional security without reducing too much the mean photon number of the pulses.

\begin{figure}[htbp]
\begin{center}
\includegraphics[width = 15 cm]{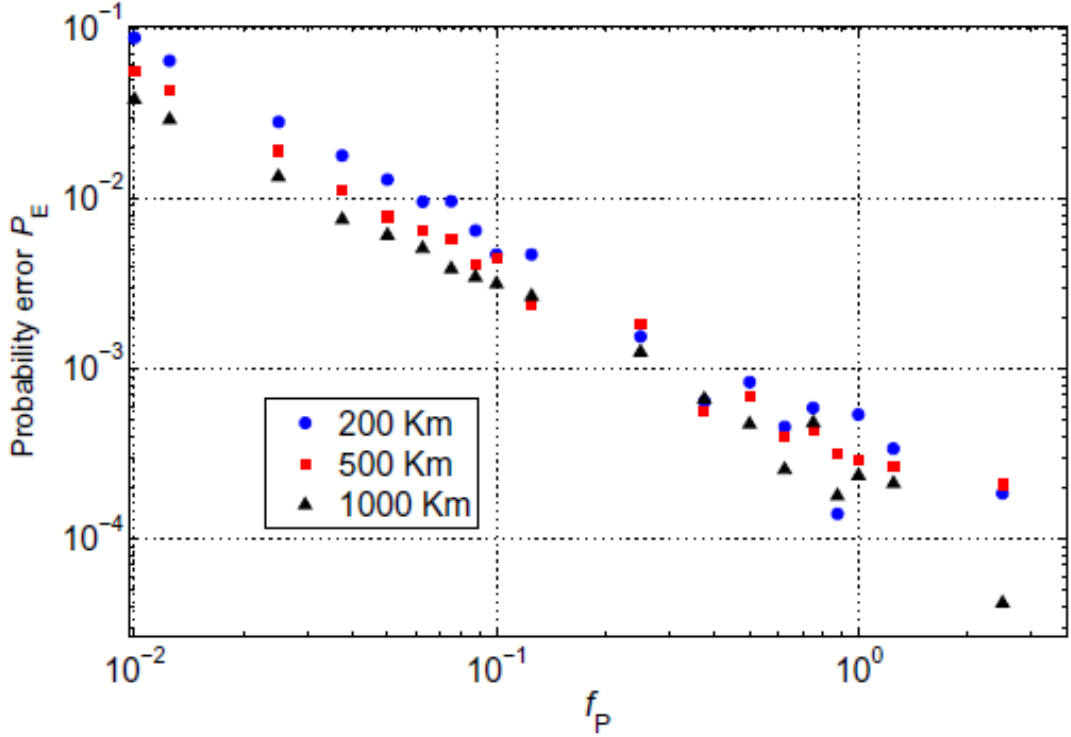}
\caption{\footnotesize Key-generation rate for the BB84 protocol using a three-level decoy-state protocol (vacuum, $\mu = 0.27$, $\mu' = 0.4$). A secure key rate of $10^{-6}$ can be obtained for an uplink to satellite orbiting at around 500 km.}
\label{figure14}
\end{center}
\end{figure}

In Fig. \ref{figure14} we show some simulations performed for a three-state decoy method, which employs the vacuum and two coherent-beam intensities $\mu = 0.27$ and $\mu' = 0.4$. Clearly there is a significant improvement in the key-generation rate, from $O (\eta^2)$ to $O(\eta)$. For a source repetition rate of $10$ MHz, in the case of an uplink to a satellite at 500 km, we would still be able to get a key generation rate of 10 bits per seconds, as compared to the value of $10^{-5}$ bits per second one would get for the BB84 protocol with no decoy states.

The main problem in the practical implementation of the decoy-state technique in a a satellite link is the unavoidable intensity fluctuations in such a link due to the fast relative motion of the communication terminals. The situation of decoy-state QKD with intensity fluctuations has been recently analyzed by Xian-Bin Wang in \cite{xianPRA07}, who showed that if the intensity-error of each pulse is random, the decoy-state protocol can work efficiently even in the case of large intensity errors.

\subsection {Entangled photons}
A detailed analysis of the conditions to violate Bell inequalities and implement a quantum key distribution experiment based on Ekert's protocol has been presented in \cite{aspelmeyerIEEE03}. As the minimum requirement they assume the SNR needed to violate a Bell-inequality \cite{FuchsPRA97}. For the case of polarization-entangled photons this necessitates a coincidence visibility of at least 71 percent, corresponding to a SNR of 6 : 1. Below that ratio it is possible to model the observed correlation with a local realistic theory, allowing unobserved eavesdropping.

\begin{figure}[htbp]
\begin{center}
\includegraphics[width = 15 cm]{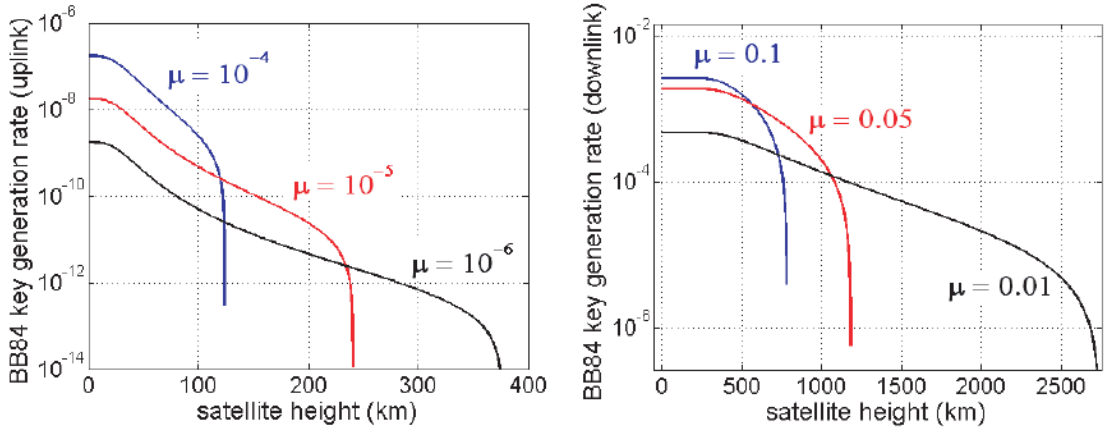}
\caption{\footnotesize SNR (in dB) for entanglement-based experiments in different configurations.}
\label{figure15}
\end{center}
\end{figure}

However, in the analysis they only consider the effect of dark counts, neglecting the effect of background noise. The rate of accidental coincidences is:
\begin{equation}
C_{acc} = N_1 N_2 \Delta t
\end{equation}
while the rate of good coincidencres is:
\begin{equation}
C = P_0 \eta_1 \eta_2
\end{equation}
where $\eta_i$ is the efficiency of the link $i$. In our simulations we use the link efficiencies and the noise values calculated in Section 2 for satellite links, while in the case of local detection we use $\eta_i \approx 0.5$ and for the noise counts just the detector dark counts ($N_i \approx 200$ counyts per second). $P_0$ is the emission rate of the entangled-photon pairs: values of the order of $10^6-10^7$ pairs per second are currently available using for examples periodically-poled nonlinear crystals.

We consider four different scenarios:
\begin{itemize}
	\item source is on the satellite, with two ground receivers (the scheme proposed for the SpaceQUEST experiment \cite{spacequest})
	\item source on the satellite with one local receiver and the other one on ground
	\item source on ground, with two satellite-based receivers
	\item source on ground with one local receiver and the other one on satellite
\end{itemize}
All simulations were performed for night-time new moon conditions. The results are shown in Fig.\ref{figure15}. It is clear that entanglement-based experiments with one photon measured locally at the source and the other one propagating either in the uplink or downlink are feasible, due to sufficient SNR (of the order of 100:1 to 1000:1). On the other hand a ground-based source with two uplinks to satellite is clearly un-feasible. The situation with a source on the satellite and two Earth-based receiving telescope is feasible, but only under some stringent requirements on the experimental parameters (telescope diameter, link distance, filtering...).

\section {Conclusions}
In this paper we discussed some aspects of the feasibility of satellite-based quantum key distribution which we believe were not yet addressed in the literature.

First of all, we discussed signal propagation through a turbulent atmosphere, refining the models presented in \cite{aspelmeyerIEEE03}, \cite{raritySat}. In particular for the uplink we analyzed the relative contribution of beam spreading and wandering, showing that the former is more important than the latter for low-altitude satellites. This makes the possibility of compensating the beam wandering with an active tip/tilt mirror not worth. Then we introduced a model for the background noise of the channel during night-time and day-time, and we discussed the signal-to-noise ratio for different configurations.

Second, we discussed the polarization properties of a satellite-based quantum channel, discussing two possible compensation techniques to the effects illustrated in \cite{bonatoOE06}. For both techniques (channel-probing at a different wavelength and time-multiplexing of signal and probe pulses at the same wavelength) we showed that the bit error rate can be kept at really low levels.

Finally we discussed the generation rate of a secure key for different configurations and for different protocols. For the standard BB84 protocol (with Poissonian-distributed source) we showed that a QKD link can be established for the downlink with a good generation rate, but not for the uplink. On the other hand, a QKD uplink can be realized with the more accurate estimate of the fraction of bits for which an eavesdropper could have complete information without introducing any disturbance, provide by the decoy-state techniques. Two points are still unclear in our opinion: the effect of the finite duration of the satellite link to the securire key generation and the possibility to implement the decoy-state technique in a channel with strong and random intensity fluctuations.
We also discussed the implementation of entanglement-based links, showing that configurations with one photon detected locally at the source and one propagating either in pulink or downlink is feasible with realistic experiemntal parameters. The situation with a source on satellite and two ground-based receivers is also feasile, but with particular care on the choice of the relevant hardware parameters.

In conclusion, we believe that satellite-based quantum key distribution is certainly feasible with present technology. We also believe that Space technology can provide a rich environment for experiments on foundational quantum mechanics and on quantum-information applications.

\section {Acknowledgements}
The authors would like to thank dr. Georg Kirchner of the Space Research Center in Graz for assistance with the KHz laser-ranging data. This research was supported by the Italian Space Agency via the Phase A feasbility study SpaceQ and by the research project QUINTET of the Department of Information Engineering, University of Padova.

\section {References}
\bibliographystyle{unsrt.bst}

\end{document}